\RequirePackage{fix-cm}
\documentclass[smallextended]{svjour3}       
\smartqed  
\usepackage{graphicx}
\begin{document}

\title{Constrained $H_\infty$ Consensus with Nonidentical Constraints
}

\titlerunning{Constrained $H_\infty$ Consensus}        

\author{Lipo Mo         \and
Yingmin Jia \and
        Yongguang Yu 
}


\institute{Lipo Mo \at
              School of Mathematics and Statistics,
        Beijing Technology and Business University, Beijing 100048, P.~R.~China \\
              \email{beihangmlp@126.com}           
          \and
           Yingmin Jia \at
             The Seventh Division,
        Beihang University, Beijing 10037, P.~R.~China
           \and
           Yongguang Yu \at
             School of Science,
        Beijing Jiaotong University, Beijing 10044, P.~R.~China
}

\date{Received: date / Accepted: date}

\maketitle

\begin{abstract}
This note considers the constrained $H_\infty$ consensus of multi-agent networks with nonidentical constraint sets. An improved distributed algorithm is adopted and a nonlinear controlled output function is defined to evaluate the effect of disturbances. Then, it is shown that the constrained $H_\infty$ consensus can be achieved if some linear matrix inequality has positive solution. Finally, the theoretical results were examined by simulation example.
\end{abstract}

\section{Introduction}

In the past decades, the consensus of multi-agent networks has received extensive attention from systems and control field. Many papers have reported the related results, such as \cite{Moreau2005,Ren2005,Xiao2006,Hong2007,Ren2008,Fu2020}. In real applications, the agents might subject to all kinds of constraints, such as position constraints \cite{Nedic2010,Lin2014}, velocity constraints \cite{Lin2017} and input constraints \cite{MoLin2018}. Most of these results, however, didn't take the external disturbances into account. There always exist disturbances in practical systems due the uncertainties of environment and the existence of model errors \cite{Mo2013,Jia2000}. To deal with the external disturbances, the concept of the $H_\infty$ consensus was introduced for first order multi-agent networks based on the robust $H_\infty$ control theory \cite{Lin2008}. Then, this method was extended to second or high order multi-agent networks \cite{Lin2009,Li2009,Mo2011,Liu2011}, where the states of all agents were assumed to be free. In some situations, we need to consider the target constrained consensus, where all agents were required to reach consensus on some constraint set, such as \cite{Shi2013}. When the target constraint was taken into account, the constrained $H_\infty$ consensus problem was studied in \cite{LinRen2017}. But, it was assumed that all constraint sets were identical. It is still unclear when the constraint sets are nonidentical. The heterogeneousness of dynamics of different agents brings us more challenges in analyzing the constrained consensus and $H_\infty$ performance, which leads to the analysis methods in \cite{LinRen2017} can not be directly used here anymore.

In this note, we are plan to consider the constrained $H_\infty$ consensus of multi-agent networks with nonidentical constraints. An improved nonlinear algorithm is employed, which was first introduced in \cite{Shi2013}, where each agent can only access the data of its own constraint set. At the same time, the controlled output is defined to evaluate the effect of the disturbances. By constructing the Lypunov functions and borrowing $H_\infty$ control theory, the LMI conditions for constrained $H_\infty$ consensus are obtained. Compare with \cite{LinRen2017}, where the constraint sets are identical, the constraint sets in this note are assumed to be nonuniform, which means the problem in \cite{LinRen2017} is the special case of this note.

{\bf Notations.} Let $\mathbf{R}^m$ be the $n$-dimension real space; Given $x \in \mathbf{R}^m$, $x^T$ denotes its transpose; $I_n$ denotes the identity matrix; Let $Y \subset \mathbf{R}^m$ be a bounded closed convex set, $y \in \mathbf{R}^m$, define $P_Y(y) = arg\min_{x\in Y} \|x - y\|$, which is the projection of $y$ on $Y$.

\section{Problem Statements}\label{sec2}

Consider a multi-agent network with $n$ agents, which can be represented by the nodes of an undirected graph $G = (V, E)$, where $V = \{1, 2, \cdots, n\}$ is the node set and $E$ is the edge set. If agents $i$ and $j$ can share information, then the edges $(i, j), (j, i) \in E$ and the weight is denoted by $a_{ij} = a_{ji} > 0$, otherwise $a_{ij} = 0$. Suppose the dynamics of agents are as follows
\begin{equation}\label{non1}
\dot{x}_i(t) = u_i(t) + w_i(t), \ \ i \in V,
\end{equation}
where $x_i(t), u_i(t) \in \mathbf{R}^m$ is the state and control input of agent $i$, $w_i(t) \in L_2 [0, + \infty)$ is the external disturbance input. Let $X_i$ be the closed convex constraint set, which can only be accessed to by agent $i$. We define the controlled output as
\begin{equation}\label{non2}
z(t) = [z_1(t)^T, z_2(t)^T]^T,
\end{equation}
where $z_1(t) = c_1 [x_1(t)^T - P_{X}(x_1(t))^T, \cdots, x_n(t)^T - P_{X}(x_n(t))^T]^T$, $z_2(t) = c_2 [x_1(t)^T$ $- \overline{x}(t)^T, \cdots, x_n(t)^T - \overline{x}(t)^T]^T$, $\overline{x}(t) = \frac{1}{n} \sum_{i=1}^n x_i(t)$, $X = \cap_{i=1}^n X_i \neq \emptyset$ and $c_1, c_2 > 0$ are two parameters.

The task of this note is to propose an effective distributed algorithm to force all agents achieve constrained $H_\infty$ consensus on $X$, i.e., when $w_i(t) = 0$ for all $i$, there exists $x^* \in X$ such that $\lim_{t\rightarrow\infty} \|x_i(t) - x^*\| = 0$; when $w_i(t) \neq 0$, the $H_\infty$ performance index $J(t) = \int_0^t [z(t)^Tz(t) - \gamma^2 w(t)^Tw(t)]dt < 0$ for all $t > 0$ and $x_i(0) = x_j(0) \in X$, where $\gamma > 0$ is the upper bound of the $L_2$ gain from disturbance to output. We first give some lemmas before our main results.

\begin{lemma}\rm\cite{tulun}
Let $G$ be a undirected connected graph, $L$ be its Laplacian and $\lambda_1 \leq \lambda_2 \leq \cdots \leq \lambda_n$ be the eigenvalues of $L$. Then $\lambda_2 > \lambda_1 = 0$.
\end{lemma}
\begin{lemma}\rm\cite{Lin2008}
Let $\Phi = 1_n - \frac{1}{n}1_n 1_n^T$. There exists a orthogonal matrix $U = [U_1\ \ \frac{1_n}{\sqrt{n}}] \in \mathbf{R}^{n\times n}$, such that $U^T \Phi U = \left[
           \begin{array}{cc}
             I_{n-1} & 0 \\
             0 & 0 \\
           \end{array}
         \right]
$ and $U^T L U = \left[
           \begin{array}{cc}
             \overline{L} & 0 \\
             0 & 0 \\
           \end{array}
         \right]$, where $1_n$ is an $n$ dimension column vector with all components being one and $\overline{L} = U_1^T L U_1$.
\end{lemma}
\begin{lemma}\rm\cite{Nedic2010}
Let $\emptyset \neq Y \subset \mathbf{R}^m$ be a closed convex set. For any $x, z \in \mathbf{R}^m$ and $y \in Y$, $\|P_Y(x) - y\|^2 \leq \|x - y\|^2 - \|P_Y(x) - x\|^2$, $[x - P_Y(x)]^T[y - x] \leq 0$ and $\|P_Y(x) - P_Y(z)\| \leq \|x - z\|$.
\end{lemma}

\section{Main Results}

To complete the target of this note, we adopt the following algorithm
\begin{equation}\label{non3}
u_i(t) = \sum_{j \in N_i} a_{ij}(x_j(t) - x_i(t)) + k_i [P_{X_i}(x_i(t)) - x_i(t)], \ \ i \in V,
\end{equation}
where $k_i > 0$ is the feedback gains and $N_i = \{j\in V | (j, i) \in E\}$ is the neighbor set of agent $i$. Thus, the closed-loop system can be rewritten as
\begin{equation}\label{non4}
\dot{x}_i(t) = \sum_{j \in N_i} a_{ij}(x_j(t) - x_i(t)) + k_i [P_{X_i}(x_i(t)) - x_i(t)] + w_i(t).
\end{equation}

\begin{theorem}\rm
Suppose the communication graph is connected. If there exists $a > \frac{\overline{k}^2}{\lambda_2^2}$, such that $\Gamma < 0$, then the constrained $H_\infty$ consensus can be achieved, where $\overline{k} = \max \{k_1, k_2, \cdots, k_n\}$, $\lambda_2$ is the smallest nonzero eigenvalue of the Laplacian of the communication graph and
\[
\Gamma = \left[
           \begin{array}{ccc}
             -(2\lambda_2 a - c_1^2) & \frac{\overline{k}}{2} & a \\
             \frac{\overline{k}}{2} & -(\lambda_2 - c_2^2) & \frac{1}{2} \\
             a & \frac{1}{2} & -\gamma^2 \\
           \end{array}
         \right].
\]
\end{theorem}
{\bf Proof.} We first consider the zero input response of system (\ref{non4}). Let $V_1(t) = \frac{1}{2} \sum_{i=1}^n \|\widetilde{x}_i(t)\|^2$, where $\widetilde{x}_i(t) = x_i(t) - P_{X}(x_i(t))$ for all $i$. Note that $[x_i(t) - P_X(x_i(t))]^T \frac{d}{dt}P_X(x_i(t)) = 0$, we have
\[
\begin{array}{lll}
\dot{V}_1(t) = \sum_{i=1}^n [x_i(t) - P_X(x_i(t))]^T \dot{x}_i(t)\\[0.2cm]
=  \sum_{i=1}^n [x_i(t) - P_X(x_i(t))]^T [\sum_{j \in N_i} a_{ij}(x_j(t) - x_i(t)) + k_i P_{X_i}(x_i(t)) - x_i(t)]\\[0.2cm]
= \sum_{i=1}^n [x_i(t) - P_X(x_i(t))]^T [\sum_{j \in N_i} a_{ij}[(x_j(t) - P_X(x_j(t)))\\[0.2cm] \ \ \ \ - (x_i(t) - P_X(x_i(t)))] + \sum_{j \in N_i}a_{ij} (P_X(x_j(t)) - P_X(x_i(t)))\\[0.2cm]\ \ \ \ \ \ + k_i [P_{X_i}(x_i(t)) - x_i(t)]].\\[0.2cm]
\end{array}
\]
Since $X$ is closed and convex, the angle between $x_i(t) - P_X(x_i(t))$ and $P_X(x_j(t)) - P_X(x_i(t))$ must lie in $[\frac{\pi}{2}, \pi]$, which implies that \[ [x_i(t) - P_X(x_i(t))]^T [P_X(x_j(t)) - P_X(x_i(t))] \leq 0.\]
If $x_i(t) \in X_i$, then $[x_i(t) - P_X(x_i(t))]^T [P_{X_i}(x_i(t)) - x_i(t)] = 0$. If $x_i(t) \notin X_i$, then the angle between $x_i(t) - P_X(x_i(t))$ and $P_{X_i}(x_i(t)) - x_i(t)$ must lie in $[\frac{\pi}{2}, \pi]$ due to the fact that $X \subset X_i$ and the convexity of $X$ and $X_i$. Hence, \[ [x_i(t) - P_X(x_i(t))]^T [P_{X_i}(x_i(t)) - x_i(t)] \leq 0.\] Therefore,
\[
\begin{array}{lll}
\dot{V}_1(t) \leq \sum_{i=1}^n [x_i(t) - P_X(x_i(t))]^T [\sum_{j \in N_i} a_{ij}[(x_j(t) - P_X(x_j(t)))\\[0.2cm]
\ \ \ \ \ \  - (x_i(t) - P_X(x_i(t)))] \\[0.2cm]
= - \widetilde{x}(t)^T (L \otimes I_m) \widetilde{x}(t) \leq - \lambda_2 \sum_{i=1}^n \|x_i(t) - P_X(x_i(t))\|^2 = - \lambda_2 V_1(t),
\end{array}
\]
where $\widetilde{x}(t) = [\widetilde{x}_1(t)^T, \cdots, \widetilde{x}_n(t)^T]^T$. Thus, $\lim_{t\rightarrow\infty} \|x_i(t) - P_X(x_i(t))\| = 0$ for all $i$. Let $y(t) = U_1^T x(t)$, $\overline{y}(t) = \frac{1}{\sqrt{n}} 1_n^T x(t)$ and $V_2(t) = \frac{1}{2} \sum_{i=1}^n \|x_i(t) - \frac{1}{n}\sum_{k=1}^n x_k(t)\|^2$. Note that $\sum_{i=1}^n [x_i(t) - \frac{1}{n} \sum_{j=1}^n x_j(t)]^T \frac{1}{n} \sum_{k=1}^n \dot{x}_k(t) = 0$ and $\sum_{i=1}^n \sum_{k=1}^n a_{ik}(x_k(t) - x_i(t)) = 0$, we have
\[
\begin{array}{lll}
\dot{V}_2(t) = \sum_{i=1}^n [x_i(t) - \frac{1}{n} \sum_{j=1}^n x_j(t)]^T [\dot{x}_i(t) - \frac{1}{n}\sum_{k=1}^n \dot{x}_k(t)]\\[0.2cm]
=\sum_{i=1}^n [x_i(t) - \frac{1}{n} \sum_{j=1}^n x_j(t)]^T [\sum_{j \in N_i} a_{ij}(x_j(t) - x_i(t))\\[0.2cm]
\ \ \ \ \ \ \  + k_i [P_{X_i}(x_i(t)) - x_i(t)]]\\[0.2cm]
= - x(t)^T (L\otimes I_m) x(t) - \frac{1}{n}\sum_{i=1}^n \sum_{j=1}^n x_j(t)^T \sum_{k=1}^n a_{ik}(x_k(t) - x_i(t))\\[0.2cm]
 \ \ \ \ + \sum_{i=1}^n (x_i(t) - \frac{1}{n}\sum_{j=1}^n x_j(t))^T k_i [P_{X_i}(x_i(t)) - x_i(t)]\\[0.2cm]
= - x(t)^T (L\otimes I_m) x(t)  +  \sum_{i=1}^n (x_i(t) - \frac{1}{n}\sum_{j=1}^n x_j(t))^T k_i [P_{X_i}(x_i(t)) - x_i(t)]\\[0.2cm]
=-y(t)^T (U_1^TLU_1\otimes I_m) y(t) + y(t)^T (U_1^T K\otimes I_m) \widetilde{x}(t),
\end{array}
\]
where $K  = {diag}\{k_1, k_2, \cdots, k_n\}$. Let $V(t) = 2a V_1(t) + V_2(t)$, $a > \frac{\overline{k}^2}{\lambda_2^2}$, then
\[
\begin{array}{lll}
\dot{V}(t) \leq -2a \lambda_2 \sum_{i=1}^n \|x_i(t) - P_X(x_i(t))\|^2 - y(t)^T (U_1^T L U_1\otimes I_m) y(t)\\[0.2cm]
 \ \ \ \ \ \ \ \ \ + y(t)^T (U_1^T K \otimes I_m) \widetilde{x}(t)\\[0.2cm]
\leq -2a\lambda_2 \sum_{i=1}^n \|x_i(t) - P_X(x_i(t))\|^2 - \lambda_2 y(t)^T y(t) + \frac{1}{2}\lambda_2 y(t)^T y(t)\\[0.2cm]
 \ \ \ \ \ \ \ \ \ + \frac{2\overline{k}^2}{\lambda_2}\sum_{i=1}^n \|x_i(t) - P_X(x_i(t))\|^2\\[0.2cm]
= -(2a\lambda_2 - \frac{2\overline{k}^2}{\lambda_2})\sum_{i=1}^n \|x_i(t) - P_X(x_i(t))\|^2 - \frac{1}{2}\lambda_2 y(t)^T y(t).
\end{array}
\]
Hence, $\lim_{t\rightarrow\infty} \sum_{i=1}^n \|x_i(t) - P_X(x_i(t))\|^2 = 0$ and $\lim_{t\rightarrow\infty} y(t)^Ty(t) = 0$. Since $x(t) = U [y(t)^T, \overline{y}(t)^T]^T = [U_1 \ \frac{1}{\sqrt{n}} 1_n] [y(t)^T, \overline{y}(t)^T]^T$, we have $$\lim_{t\rightarrow\infty} \|x_i(t) - x_j(t)\| = 0,$$ i.e., there exists $x^* \in \mathbf{R}^m$ such that $\lim_{t\rightarrow\infty} \|x_i(t) - x^*\| = 0$. Hence, $\|x^* - P_X(x^*)\| = \lim_{t\rightarrow\infty} \|x_i(t) - P_X(x_i(t))\| = 0$, which implies that $x^* = P_X(x^*) \in X$.

Next, let us consider the $H_\infty$ performance of system (\ref{non4}) with controlled output (\ref{non2}). Based on the $H_\infty$ control theory, we assume $x_i(0) = x_j(0) \in X$ for all $i$ and $j$. Here, we have
\[
\begin{array}{lll}
\dot{V}(t) \leq -2a\lambda_2 \sum_{i=1}^n \|x_i(t) - P_X(x_i(t))\|^2 + 2a \sum_{i=1}^n [x_i(t) - P_X(x_i(t))]^T w_i(t)\\[0.2cm]
- \lambda_2 y(t)^T y(t) + y(t)^T (U_1K \otimes I_m) \widetilde{x}(t) + \sum_{i=1}^n (x_i(t) - \frac{1}{n}\sum_{j=1}^n x_j(t))^T w_i(t)\\[0.2cm]
= -2 \lambda_2 a \|\widetilde{x}(t)\|^2 + 2 a \widetilde{x}(t)^T w(t) - \lambda_2 \|y(t)\|^2 + y(t)^T (U_1K\otimes I_m) \widetilde{x}(t) \\[0.2cm]
\ \ \ \ \ + x(t)^T (\Phi\otimes I_m) w(t)\\[0.2cm]
= -2 \lambda_2 a \|\widetilde{x}(t)\|^2 + 2 a \widetilde{x}(t)^T w(t) - \lambda_2 \|y(t)\|^2 + y(t)^T (U_1K\otimes I_m) \widetilde{x}(t) \\[0.2cm]
\ \ \ \ \  + y(t)^T (U_1^T\otimes I_m) w(t)\\[0.2cm]
\leq - 2 \lambda_2 a \|\widetilde{x}(t)\|^2 + 2 a \|\widetilde{x}(t)\| \|w(t)\| - \lambda_2 \|y(t)\|^2 + \overline{k} \|y(t)\| \|\widetilde{x}(t)\| \\[0.2cm]
\ \ \ \ \  + \|y(t)\| \|w(t)\|.
\end{array}
\]
Note that $z(t)^T z(t) = z_1(t)^T z_1(t) + z_2(t)^Tz_2(t) = c_1 \|\widetilde{x}(t)\|^2 + c_2 \|\Phi x(t)\|^2 \leq c_1 \|\widetilde{x}(t)\|^2 + c_2 \|y(t)\|^2$, $\Gamma < 0$, $V(0) = 0$ and $V(t) \geq 0$ for all $t > 0$, we have
\[
\begin{array}{lll}
J(t) = \int_0^t [z(s)^Tz(s) - \gamma^2 w(s)^Tw(s)] ds \\[0.2cm]
= \int_0^t [z(s)^Tz(s) - \gamma^2 w(s)^Tw(s) + \dot{V}(s)] ds - V(t) \\[0.2cm]
\leq \int_0^t \xi(s)^T \Gamma \xi(s) ds < 0,
\end{array}
\]
for all $t > 0$, where $\xi(t) = [\|\widetilde{x}(y)\|, \|y(t)\|, \|w(t)\|]^T$. Therefore, the $H_\infty$ performance index is also satisfied. This completes the proof of Theorem 1.

\begin{remark}\rm
By Schur Lemma \cite{Jia}, $\Gamma < 0$ is equivalent to $-(2a\lambda_2 - c_1^2) + \frac{a^2}{\gamma^2} + (\frac{\overline{k}}{2} + \frac{a}{2\gamma^2})^2 [\lambda_2 - c_2 - \frac{1}{4\gamma^2}]^{-1} < 0$, which always have solution on $a$ when $\lambda_2$ is large enough, which can be implemented by selecting large enough weights $a_{ij}$ of communication graph.
\end{remark}

\section{Conclusions}\label{sec6}
In this technical note, we considered the constrained $H_\infty$ consensus of multi-agent networks with nonidentical target constraints. A linear matrix inequality condition was obtained for constrained $H_\infty$ consensus by Lyapunov function and $H_\infty$ control method. Finally, the simulation was provided to justify the effectiveness of our theoretical results. In the future, we will consider the situations of time-delays and switching topologies.

%

\begin{acknowledgements}
This work is supported by National Natural Science
Foundation (NNSF) of China (Grant Nos. 61973329) and the Beijing Natural Science Foundation (Grant Nos. Z180005). 
\end{acknowledgements}

%
%



\end{document}